\documentclass[a4paper,fleqn]{cas-dc}

\usepackage[authoryear]{natbib}
\usepackage{aas_macros}

\usepackage{subcaption}

\def\tsc#1{\csdef{#1}{\textsc{\lowercase{#1}}\xspace}}
\tsc{WGM}
\tsc{QE}

\def\Msun{\mbox{M$_\odot$}}

\def\ML{\mbox{$M/L$}}

\def\mst{\mbox{$M_{\star}$}}

\def\lsim{\mathrel{\rlap{\lower3.5pt\hbox{\hskip0.5pt$\sim$}}
    \raise0.5pt\hbox{$<$}}}                
\def\gsim{~\rlap{$>$}{\lower 1.0ex\hbox{$\sim$}}}

\def\Fig{\mbox{Fig.~}}

\def\Sec{\mbox{Section~}}

\def\Re{\mbox{$R_{\rm e}$}}

\newcommand{\ion}[2]{#1\,\textsc{#2}}

\newcommand{\camels}{$\textsc{camels}$}

\begin{document}
\let\WriteBookmarks\relax
\def\floatpagepagefraction{1}
\def\textpagefraction{.001}

\shorttitle{Golden mass in galaxy evolution}    

\shortauthors{C.~Tortora}  

\title [mode = title]{Understanding galaxies and the origin of the golden mass through the study of scaling relations across cosmic history}  

\tnotemark[1] 

\tnotetext[1]{} 

%

\author[1]{C.~Tortora}

\cormark[1]

\fnmark[1]

\ead{crescenzo.tortora@inaf.it}

\ead[url]{www.crescenzotortora.com}

\credit{Conceptualization of this study, Methodology, Software - Original Draft, Data interpretation from related works}

\affiliation[label1]{organization={INAF -- Osservatorio Astronomico di Capodimonte},
            addressline={Salita Moiariello 16}, 
            city={Napoli},
            postcode={I-80131}, 
            country={Italy}}

\author[2]{A.~Gallazzi}
\credit{}
\affiliation[label2]{organization={INAF -- Osservatorio Astrofisico di Arcetri},
            addressline={Largo Enrico Fermi 5}, 
            city={Firenze},
            postcode={I-50125}, 
            country={Italy}}

\author[3]{A.~Gargiulo}
\credit{}
\affiliation[label3]{organization={INAF -- Istituto di Astrofisica Spaziale e Fisica Cosmica},
            addressline={ Via A. Corti 12}, 
            city={Milano},
            postcode={I-20133}, 
            country={Italy}}

\author[1]{F.~La~Barbera}
\credit{}

\author[4]{N.~R.~Napolitano}
\credit{}
\affiliation[label4]{organization={Department of Physics E. Pancini, University Federico II},
            addressline={Via Cinthia 6}, 
            city={Napoli},
            postcode={I-80126}, 
            country={Italy}}

\author[1]{R.~Ragusa}
\credit{}


\author[2]{S.~Zibetti}
\credit{}

\cortext[1]{Crescenzo Tortora}

\fntext[1]{}


\begin{abstract}
The total mass distribution in galaxies strongly influences the baryonic cycle and star formation (SF), both directly, through gas availability and feedback from supernovae and AGN, and indirectly via the depth of the dark matter (DM) potential well. The star formation efficiency, defined as the stellar-to-total mass ratio normalized by the cosmological baryon fraction, peaks at a characteristic stellar mass of $5 \times 10^{10} \Msun$, often termed the ``golden mass''. At this threshold, various feedback mechanisms minimally affect SF. This critical mass scale also appears in other galaxy scaling relations. Among the many known trends, the correlations between stellar population gradients, DM fraction, and total mass density slope as a function of stellar mass all show pronounced minima at the golden mass. What physical processes shape these observed relations? And how, why, and when does the golden mass emerge during galaxy evolution? Early cosmic epochs ($z \gtrsim 1.5$–2) remain largely unexplored, yet they probe the formative stages of galaxy mass assembly, the era when the golden mass scale first emerges, and provide the full redshift baseline needed to constrain the buildup of stellar mass and the efficiency of gas-to-star conversion in massive galaxies. By combining high resolution, multiplexing, and a wide field of view, SHARP on the ELT can address these open questions, delivering sharper and deeper data than  James Webb Space Telescope ({\it JWST}). In this contribution to the SHARP White Paper, we present a scientific case that exploits these unique characteristics. With its two observing modes, SHARP will enable measurements of (a) stellar population gradients, (b) DM fractions, and (c) total mass density slopes. In particular, with the mIFU mode, SHARP will be able to obtain spatially-resolved stellar populations and dynamical maps for galaxies at $z \gsim 1.5$. Reconstructing the scaling relations of these observables as a function of stellar mass for early- and late-type systems would allow us to trace these scaling relations and the emergence of the golden mass as a function of redshift, and to reconstruct the impact of galaxy processes (such as mergers, SN, and AGN feedback) by comparing them with cosmological simulations.
\end{abstract}





\maketitle





\section{Introduction}

Galaxy evolution arises from the interplay between two broad classes of processes, commonly referred to as ``nature and nurture''. The former encompasses internal, self-regulated mechanisms such as supernova (SN) and Active Galactic Nuclei (AGN) feedback, which govern the baryon cycle. The latter includes external, environment-driven effects: interactions among galaxies and between galaxies and the intracluster or intergalactic medium, shock heating and cold streams \citep{dek_birn06, Waterval+25}. All these processes operate within a hierarchical framework, where galaxy mergers play a fundamental role in driving mass assembly and shaping galaxy properties across cosmic time. \textit{Some broad and key fundamental questions are: how do these mechanisms combine to drive the formation and evolution of massive galaxies throughout cosmic history, and to what extent can nature be disentangled from nurture?}

Galaxy scaling relations (e.g. those linking structural parameters, star-formation rate (SFR), colours, stellar population properties (age and metallicity) and their gradients, velocity dispersion, central dark matter (DM) fraction, DM density, and mass density slope to stellar mass and many others) encode key information about the physical processes that drive galaxy formation and evolution \citep[e.g.,][]{FJ76,Kauffmann+03,HB09_FP,Gallazzi+05,Cappellari+06,Koopmans+09,Tortora+09,Tortora+10CG,Tortora+14_DMslope,Tortora+18_KiDS_DMevol,Tortora+19_LTGs_DM_and_slopes,NRT10,Moster+10,Treu+10,Whitaker+14,Speagle+14,Daddi+22_main_sequence,Popesso+23}. These correlations are complex, showing monotonic or non-monotonic trends that can evolve with redshift and environment.

A crucial revelation from these scaling relations in the local Universe is the existence of a characteristic threshold known as the ``golden mass'', located at a stellar mass of roughly $\mst \sim 5 \times 10^{10}\, \Msun$ \citep[e.g.,][]{dek_birn06, Tortora+25_CASCOIII}. At this critical juncture, global star-formation efficiency reaches its peak. Furthermore, several fundamental scaling relations exhibit pronounced minima or turnovers at this exact scale. Specifically, as a function of stellar mass, (a) stellar population gradients, (b) central DM fractions, and (c) total mass density slopes all reach their most extreme (steepest or minimum) values at the golden mass. Below this scale, galaxy evolution is predominantly regulated by SN feedback; above it, AGN feedback and hierarchical mergers dominate. 

While the golden mass and its associated scaling relations are well-established at low redshift, their origins remain largely unexplored in the early Universe. Cosmic epochs at $z \gtrsim 1.5$--$2$ represent the formative stages of galaxy mass assembly, marking the era when the golden mass scale is theoretically expected to first emerge. Tracing these specific correlations at high redshift is essential to constrain the buildup of stellar mass and the efficiency of gas-to-star conversion. However, doing so requires observing the central kinematics and spatially resolved stellar populations of thousands of galaxies. Current observational facilities and surveys, including \textit{JWST}, lack the required combination of multiplexing, wide field of view, and extreme spatial resolution necessary to build statistically robust samples of these precise observables beyond $z \sim 1.5$.

In this context, this contribution to the Single High Angular Resolution and Precision (SHARP) White Book presents the scientific case for using the upcoming SHARP spectrograph on the Extremely Large Telescope (ELT) to overcome this observational bottleneck. By leveraging its two complementary observing modes, i.e. the multi-object spectrograph NEXUS and the multi-IFU spectrograph VESPER, SHARP will deliver integrated and spatially resolved spectroscopy for thousands of early- and late-type galaxies at $z > 1.5$. Our primary objective is to reconstruct the scaling relations of stellar population gradients, DM fractions, and total mass density slopes as a function of redshift. By doing so, we will trace the emergence of the golden mass across cosmic time and directly constrain the impact of fundamental galaxy processes by comparing these unprecedented observations with cosmological simulations.

\section{Scientific rationale and open questions}

Galaxy evolution is governed by a number of characteristic mass scales that regulate the relative importance of different physical processes. Among these, the golden mass represents a key transition, where the balance between stellar and AGN feedback is thought to change dramatically. In this section, we review the current observational and theoretical picture of the golden mass.

\subsection{Golden mass from observed scaling relations}\label{subsec:golden_mass_obs}

The golden mass is found at approximately $M_{\rm halo} \sim 10^{12}\, \Msun$ and $\mst \sim 5 \times 10^{10}\, \Msun$ \citep{dek_birn06,Tortora+25_CASCOIII}. This mass scale corresponds to the peak of the global star-formation efficiency, or equivalently to the bend in the stellar-to-halo mass relation \citep[SHMR;][]{Moster+10, Behroozi+13, Behroozi+19, Girelli+20}, one of the earliest and most extensively studied relations to reveal the existence of such a characteristic scale. Below and above this mass scale, feedback mechanisms (e.g., SN-driven winds at low mass and AGN feedback) become increasingly effective at suppressing star formation. Moreover, \cite{dek_birn06} proposed that one of the primary drivers of these trends is the transition from cold gas accretion to virial shock heating at a critical halo mass of $\sim 10^{11.7} \, \Msun$. This transition, together with the presence of cold streams at high redshift, the regulation of star formation by feedback processes, and the effects of gravitational clustering, plays a key role in shaping galaxy evolution \citep{Daddi+22_main_sequence, Daddi+22_cold_hot}.

The golden mass also appears as a minimum in the correlation of the central DM fraction \citep{Wolf+10, Toloba+11_I, Cappellari+13_ATLAS3D_XX, TLBN16_IMF_dwarfs, Tortora+19_LTGs_DM_and_slopes, Aquino-Ortiz+18, Lovell+18_Illustris, Busillo+23_CASCO-I, Busillo+25_CASCO-II} and the total mass density slope \citep{Tortora+19_LTGs_DM_and_slopes} with stellar mass. This mass scale has been identified in other ways, e.g. through structural and stellar population analyses, such as the bending in the size–mass relation \citep{Shen+03, HB09_curv, Tortora+09, Saracco+17, Roy+18, Nedkova+21,vanderWel+24}, in the correlations between stellar population properties and mass \citep{Kauffmann+03, Gallazzi+05, Zibetti_Gallazzi2022,Mattolini+25}, and as a minimum in optical colour, metallicity, and stellar \ML\ gradients as a function of mass \citep{Kuntschner+10, Spolaor+10, Tortora+10CG, Tortora+11MtoLgrad,Goddard+17_MANGA, Liao_Cooper2023}. Taken together, this extensive body of empirical evidence converges on a golden stellar mass of approximately $1\text{–}5 \times 10^{10}\, \Msun$ at $z \sim 0$.

While such scaling relations have been extensively explored at low redshift, only in the last two decades has the higher-$z$ universe begun to be probed more systematically, with galaxy samples growing in size. However, the analysis of less massive galaxies remains challenging due to observational limitations. The most studied correlation with redshift is the SHMR, typically inferred through halo abundance matching, which shows that the bend in the relation shifts to higher masses at earlier cosmic epochs, reaching values of $\sim 10^{11}\, \rm M_\odot$ at $z \sim 3-4$ \citep[e.g.,][]{Girelli+20}. Other correlations, such as the size–mass relation, also exhibit a bending at this characteristic mass that increases with redshift \citep{Nedkova+21}. Several studies further report a bending in the main sequence of star-forming galaxies, increasing with redshift \citep{Daddi+22_main_sequence, Popesso+23, Stephenson+24, Enia+25_Euclid}. However, \cite{Tortora+25_CASCOIII} suggest that this latter feature may correspond to a different (lower-)mass scale, i.e. the "gas-richness" mass scale in \cite{Kannappan04}, which separate accretion-dominated from processing-dominated galaxies. 

However, among these various scaling relations, we have identified a few that clearly can probe the differential role of feedback processes and galaxy mergers and reveal the emergence of the golden mass as a clear minimum, i.e., a) stellar population gradients \citep{Tortora+10CG,Tortora+11MtoLgrad,Liao_Cooper2023}, b) DM fraction, and c) mass density slope \citep{Sonnenfeld+13_SL2S_IV,Dutton_Treu14,Tortora+14_DMslope,TLBN16_IMF_dwarfs,Tortora+18_KiDS_DMevol,Tortora+19_LTGs_DM_and_slopes,Mukherjee22_SEAGLEIII,Busillo+23_CASCO-I,Busillo+25_CASCO-II,Tortora+25_CASCOIII} versus stellar mass. We intend to examine them in more detail here:

\begin{itemize}
\item \textbf{Stellar population gradients.}
Colour and metallicity gradients trace the differential impact of physical processes that can affect either the central regions or the outskirts of galaxies. At $z \sim 0$ these gradients are typically shallow at low masses ($\mst \sim 10^{9}\, \rm \Msun$), reach their steepest values around the golden mass ($\mst \sim 5 \times 10^{10}\, \rm \Msun$), and become shallower again at higher masses \citep{Kuntschner+06, Spolaor+09, Tortora+10CG,Zibetti+20}. Their interpretation generally invokes different physical mechanisms across the mass spectrum: in low-mass galaxies, colour and metallicity gradients are typically attributed to early monolithic collapse episodes, during which rapid star formation and subsequent SN feedback drive metal-poor gas to larger radii while enriching the central regions. At the high-mass end, gradients are instead shaped by the combined action of AGN feedback, which can quench or regulate star formation, and hierarchical assembly through mergers. Major mergers tend to flatten pre-existing gradients by mixing stars efficiently. See \cite{Tortora+10CG} and references therein. The role of environment has also been investigated: while age and metallicity gradients generally do not depend on environment \citep{TNgenv12,Goddard+17_gradients}, for massive \citep{LaBarbera+05,LaBarbera+11_CG} or central galaxies \citep{TNgenv12} the gradients tend to be shallower in dense environments.

\item \textbf{DM fractions.}
Measurements of the DM fraction provide additional constraints on galaxy formation and mass assembly. The DM fraction within the half-light radius is highest in low-mass galaxies ($\mst \sim 10^{9}\, \rm \Msun$), decreases toward the golden mass, and increases again at higher masses \citep{TLBN16_IMF_dwarfs,Tortora+19_LTGs_DM_and_slopes,Sharma+25}. This trend is driven by a combination of size evolution, given the size–mass relation, and the shape of the SHMR on global scales \citep{Tortora+09,NRT10}. In massive early-type galaxies, some dependence on the environment has been observed \citep{SPIDER-VI,Nigoche-Netro+19}.
\item {\bf Mass density slopes.} Even more informative than the central DM fraction, the total mass density slope captures the spatially resolved impact of physical processes: it is shallow in dwarf galaxies \citep[e.g.][]{deBlok10}, reaches its steepest values around the golden mass, and becomes shallower again in the most massive systems (see \citealt{Tortora+19_LTGs_DM_and_slopes} for more details). In the most massive early-type galaxies, these findings are consistent with strong lensing studies, which indicate approximately isothermal mass profiles \citep[e.g.,][]{Gavazzi+07_SLACSIV}. The physical interpretation explaining the trends in both mass density slopes and DM fraction is consistent with that given for the colour gradients (see \citealt{Tortora+19_LTGs_DM_and_slopes} and references therein).
\end{itemize}

These dichotomic trends with stellar mass, even if with some differences, are shared by different types of galaxies \citep{Tortora+10CG,Tortora+19_LTGs_DM_and_slopes}. 

Summarizing, these correlations encode the direct impact of physical processes on (a) stellar mass assembly, as traced by stellar population gradients, and (b) the assembly of both stellar and DM components, as inferred from measurements of stellar and DM masses and of the DM and total mass density slopes. They connect directly baryons with DM.

\begin{figure*}
\centering
\begin{subfigure}{0.46\textwidth}
\caption{$z=0$}
\includegraphics[width=\textwidth]{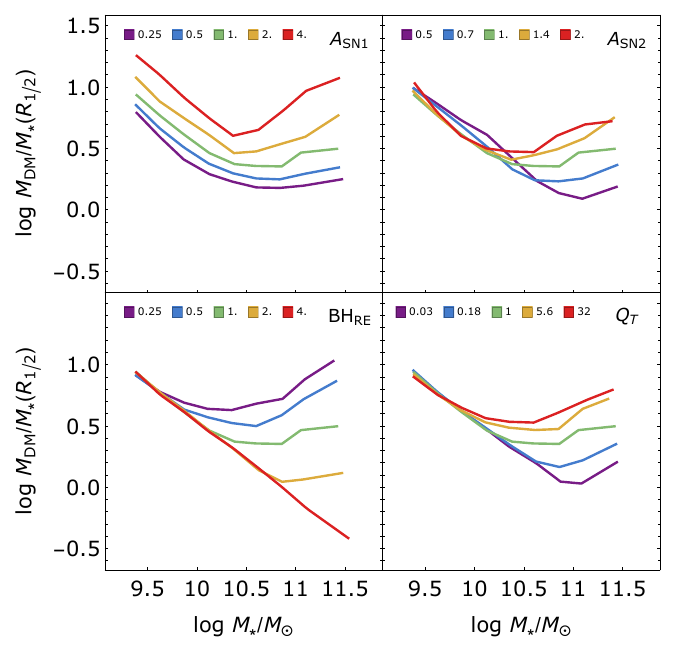}
\end{subfigure}
\hspace{0.5cm}
\begin{subfigure}{0.46\textwidth}
\caption{$z=3$}
\includegraphics[width=\textwidth]{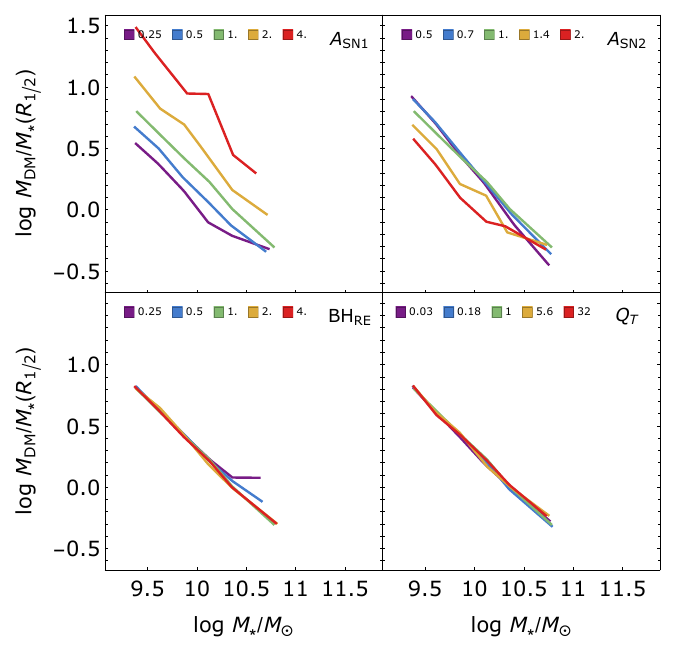}
\end{subfigure}
\caption{DM-to-stellar mass ratio within the half-mass radius $M_{\rm DM,1/2}/M_{\star,1/2}$ plotted against $M_{\star}$ from \camels\ simulations at $z=0$ (panel a) and $z=3$ (panel b). The dependence on four of the most impactful parameters is analysed: normalization factor for the energy in galactic winds per unit star formation $A_{\rm SN1}$ (top-left in each group of panels), normalization factor for the galactic wind speed $A_{\rm SN2}$ (top-right), the fraction of the accretion rest mass that is released in the accretion process $BH_{\rm RE}$ (bottom-left) and the Eddington ratio, that serves as the threshold between the low-accretion and high-accretion states of AGN feedback $Q_{\rm T}$ (bottom-right). The parameter values are normalized to those of the reference simulation  (green curve), which is calibrated to the IllustrisTNG model. The panels on the left are extracted from \cite{Tortora+25_CASCOIII}, while the panels on the right are new, although they resemble what was shown at $z=2$ in the same paper.}\label{fig:figure1}
\end{figure*}

\begin{figure}
\centering
{\includegraphics[width=0.49\textwidth]{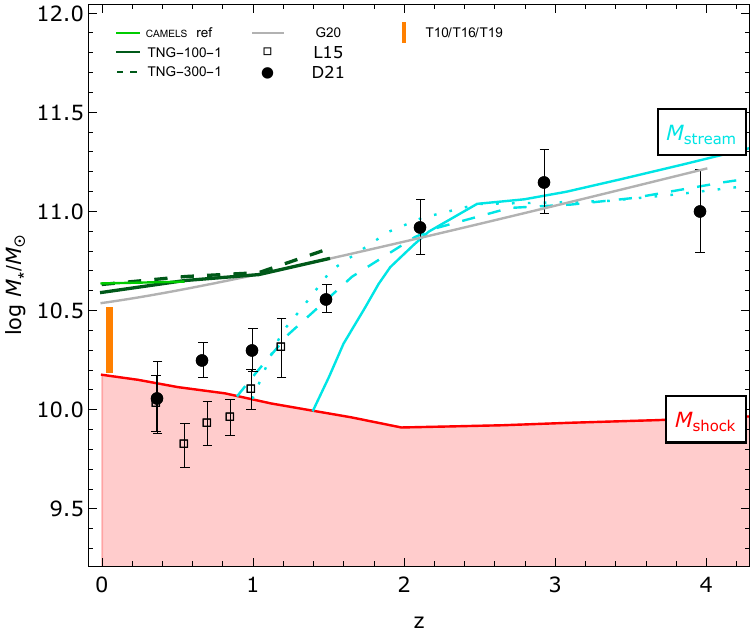}}
\caption{Golden mass determined from the $M_{\rm DM,1/2}/M_{\star,1/2} - M_{\rm \star}$ relation assuming the reference \camels\ simulation (green line), TNG100-1 (dark green line) and TNG300-1 (dashed dark green line) simulations. From the theoretical point of view, we also show the shock heating stellar mass ($M_{\rm shock}$) as a red line and the stream mass ($M_{\rm stream}$) as solid cyan line \citep{dek_birn06}, together with its two revised forms (dashed and dotted cyan lines) discussed in \cite{Daddi+22_main_sequence}. From the observational point of view, we plot the range of golden masses in the local Universe from different observables \citep{Tortora+10CG,TLBN16_IMF_dwarfs,Tortora+19_LTGs_DM_and_slopes}. Then, we also plot the stellar mass value of the bending in the stellar-to-halo mass relation determined from using halo-abundance matching in \citet[][grey solid line]{Girelli+20}, and the bending mass in the main sequence of star-forming galaxies (L15: \citealt{Lee+15}; D21: \citealt{delvecchio+21}). These figures are adapted from \cite{Tortora+25_CASCOIII}.}\label{fig:figure2}
\end{figure}

\subsection{Golden mass in cosmological simulations}\label{subsec:golden_mass_sims}

From a theoretical perspective, cosmological simulations such as EAGLE \citep{Crain+15_EAGLE} and IllustrisTNG \citep{Pillepich+18} have provided insights into the impact of physical processes on galaxy scaling relations. Simulation-based studies have also highlighted that the golden mass marks a fundamental transition in galaxy mass assembly: below this threshold, stellar mass is predominantly built through in-situ star formation, whereas above it, ex-situ accretion via mergers becomes the dominant growth channel \citep[e.g.,][]{Tacchella+19, Cannarozzo+23}. More recently, the \camels\ cosmological simulations \citep{Villaescusa-Navarro2021, Villaescusa-Navarro2023, Ni+23} have offered further understanding of the origin of scaling relations involving DM and the golden mass \citep{Tortora+25_CASCOIII}. These suites of simulations comprise thousands of runs in which cosmological parameters as well as SN and AGN feedback strengths are varied, and which adopt different subgrid physics models.

This diversity allows a broad range of applications, including detailed comparisons with observations. In the first two papers of the ``Cosmological and Astrophysical parameters from Cosmological simulations and Observations'' (CASCO) series, \cite{Busillo+23_CASCO-I} and \cite{Busillo+25_CASCO-II}, we developed a statistical method to determine, from a selection of \textsc{camels} cosmological simulations with different cosmological and feedback parameters, which simulation best reproduces observed galaxy scaling relations, focusing in particular on those involving DM mass. This analysis can potentially be extended to other scaling relations (e.g., stellar population gradients and mass density slopes) or exploited using machine learning approaches \citep[e.g.,][]{vonMarttens+22,Wu+24}.

In particular, a targeted analysis of the emergence of the golden mass in scaling relations involving the DM fraction (hereafter referred to as the DM-to-stellar mass ratio) and stellar mass \citep{Tortora+25_CASCOIII} reproduces the observed U-shaped trend of the DM-to-stellar mass ratio as a function of \mst, with a minimum at $\mst \sim 5 \times 10^{10}\, \Msun$. The \camels\ simulations show that varying the strength of SN and AGN feedback changes both the slopes and normalizations of the correlation at low and/or high masses, as well as the position of the golden mass (see the correlations at $z=0$ in the left panel of \Fig\ref{fig:figure1}). 

SN winds primarily affect the stellar mass assembly, with stronger winds leading to less massive galaxies.
In fact, stronger SN winds, i.e., winds with higher energy (parameterized by $A_{\rm SN1}$, top-left in the left panel of the figure) and higher velocity ($A_{\rm SN2}$, top-right in the same panel), shift the golden mass to lower values and steepen the high-mass end of the correlation; the two parameters impact low- and high-mass galaxies differently. 

Changes in AGN feedback primarily affect the high-mass slope of the correlation and also modify the location of the golden mass. Among the simulated parameters in the \camels\ suites, we highlight two that have the largest impact on the golden mass: the BH radiative efficiency, $BH_{\rm RE}$, i.e., the fraction of the accreted rest mass released during accretion, and the Quasar threshold, $Q_{\rm T}$, i.e., the Eddington ratio, the threshold separating low- and high-accretion states of AGN feedback. Larger values of these parameters correspond to higher and lower golden mass values, respectively.

Although this is not shown in the figure, the simulations indicate that the golden mass emerges earlier in cosmic time when feedback is stronger (see Figs. 2 and 3 in \citealt{Tortora+25_CASCOIII}). In most of the parameter space explored by \camels\ simulations, including the original IllustrisTNG runs, the U-shaped minimum corresponding to the golden mass is clearly visible up to $z \sim 1$--$1.5$, but it appears to flatten or disappear at higher redshifts, as illustrated by the $z=3$ snapshot in the right panel of \Fig\ref{fig:figure1}, where it is also evident that the effect of varying the AGN feedback is negligible, while SN feedback impacts the scaling relations also at large masses. Conversely, observational and empirical frameworks, such as the SHMR inferred from halo abundance matching, successfully track the evolution of this characteristic bending mass up to much higher redshifts, extending to $z \sim 4$ \citep[e.g.,][]{Girelli+20}. This discrepancy underscores a major open question: it remains fundamentally unknown whether the apparent disappearance of the golden mass minimum in high-redshift simulations represents a true physical transition in the early Universe or is merely an artifact of current numerical limitations, such as restricted box volumes failing to capture rare massive systems or resolution effects in subgrid feedback recipes. Pushing observations up to $z \sim 4$ with SHARP is therefore fundamentally required to resolve this theoretical bottleneck, providing the definitive empirical baseline needed to determine exactly when and how the golden mass scale was established during the earliest epochs of cosmic assembly.

Exploring the relative roles of SN and AGN feedback versus other scenarios involving virial shocks and cold streams as the dominant processes, we find that SN and AGN feedback parameters have a primary impact on the emergence of the golden mass. In fact, when their strength is reduced to the minimum, the minimum in the $M_{\rm DM,1/2}/M_{\star,1/2}$ -- $M_{\star}$ correlation, even at $z \sim 0$, tends to disappear. This indicates that, on their own, shock heating and cold stream processes alone may not be sufficient to produce the minima seen in our scaling relations, at least within the \camels\ framework using the Illustris TNG subgrid model. When compared with observed relations spanning a wide range of masses and redshifts, these simulations will constrain the strength of feedback parameters \citep[e.g.,][]{Busillo+23_CASCO-I, Busillo+25_CASCO-II}. A detailed analysis of the role of the environment has not been performed, but it could deserve some attention in the near future.

In \cite{Tortora+25_CASCOIII}, we also compared the golden mass derived from the DM-to-stellar mass relation  using some reference simulations with observational data and other theoretical expectations, which we reproduce here in \Fig\ref{fig:figure2}. The golden mass in different TNG simulation suites, including the reference \camels\ simulation based on the TNG subgrid, shows a slight increase up to $z \sim 1.5$. At $z \sim 0$, these values approximately agree with some local estimates of the golden mass obtained using the correlations identified in this paper (the correlations between colour and stellar population gradients, DM fraction, and mass density slope with stellar mass, \citealt{Tortora+10CG, TLBN16_IMF_dwarfs, Tortora+19_LTGs_DM_and_slopes}) and with the SHMR from halo abundance matching \citep{Girelli+20}. These results are also compared with the evolution of the bending mass from the main-sequence relation of star-forming galaxies \citep{Lee+15,delvecchio+21}, and with the expected shock-heating mass (i.e., the mass above which hot shocks in the halo prevent further gas accretion) and the stream mass (i.e., the mass below which cold streams can still sustain star formation). Above these two mass scales (cyan lines at $z > 1.5$ and the red line at lower redshift), galaxies are largely prevented from forming new stars \citep{dek_birn06}.

In particular, both the SHMR results from \cite{Girelli+20} and the bending of the main sequence for star-forming galaxies in \cite{Daddi+22_main_sequence} follow the trend of the stream mass at high redshift, reaching values of $\gsim 10^{11}\, \rm \Msun$  (see Figure \ref{fig:figure2}), diverging only at $z \lsim 1.5$. Therefore, we speculate that at $z \gtrsim 1.5$–2, cold streams can contribute to gas accretion \citep{Daddi+22_cold_hot}, and that a single characteristic (stream) mass, coinciding with both the bending of the stellar-to-halo mass relation \citep{Girelli+20} and the turnover in the main sequence of star-forming galaxies \citep{Daddi+22_main_sequence}, regulates galaxy evolution. At later epochs, as shown in \Fig\ref{fig:figure2}, this characteristic mass appears to bifurcate into two distinct scales: a low-mass, gas-richness scale associated with gas availability \citep{Kannappan04,Hunt+20_MAGMAII} and following the bending mass of the star-forming main sequence (converging to $\sim 5 \times 10^{9}\, \rm \Msun$ at $z\sim 0$), and a higher-mass golden mass, which governs star formation efficiency and quenching ($\sim 5 \times 10^{10}\, \rm \Msun$ at $z\sim 0$), i.e. the golden mass.

Future improvements will benefit from larger-volume simulations and higher resolution, which would increase statistics at large masses and high redshifts and reduce biases caused by limited resolution \citep{Tortora+25_CASCOIII}. In particular, this will be useful to better trace the evolution of the golden mass at $z \gtrsim 1.5-2$, and to determine whether its apparent disappearance in \camels\ and TNG simulations is due to observational limitations or to physical reasons. Within the CASCO project, future analyses will also be devoted to exploring other scaling relations, as well as, the correlation between stellar population gradients and mass density slopes as a function of mass, and also disentangle the effect of feedback phenomena from the environmental effects.

\section{Pushing observations at larger redshifts}

To contextualize our approach, it is necessary to establish a comparison between our chosen scaling relations and competing observables like the SHMR. What does each observable constrain? The SHMR provides a fundamental global perspective on the cosmological star-formation efficiency, successfully identifying the golden mass as a turnover point. However, deriving the SHMR relies often on either halo abundance matching \citep[e.g.,][]{Moster+10}, which intrinsically depends on theoretical priors from DM-only N-body simulations, or on direct halo mass measurements via weak lensing and satellite kinematics. These latter techniques trace the extreme galaxy outskirts, where the signal-to-noise (S/N) drop dramatically, and are severely plagued by observational systematics at high redshift.

In contrast, we emphasize that the three correlations introduced in \Sec\ref{subsec:golden_mass_obs} are empirically more direct and robust for tracking the internal physics of galaxy assembly. First, despite being subject to specific observational degeneracies (which SHARP is designed to break, as we will discuss in the next sections), these observables are purely empirical and completely disconnected from cosmological N-body priors. Second, they exclusively probe the central regions of galaxies (typically within $1$--$3\,\Re$). In these central regimes, observational systematics are kept strictly under control, the S/N is maximal, and the physical signatures of internal mechanisms, such as central AGN feedback or SN-driven metal outflows, are most directly imprinted. Ultimately, while both the SHMR and our proposed observables support the existence of the golden mass scale at low redshift, our internal scaling relations provide complementary diagnostics and are uniquely suited to disentangle the internal baryonic processes that drive galaxy evolution at $z \gtrsim 1.5$.

While these correlations have been extensively studied locally (see \Sec\ref{subsec:golden_mass_obs}), their investigation as a function of redshift has been only partial. Past studies were often limited to small samples, in particular the most massive galaxies up to $z \sim 1-2$, although recent \textit{JWST} observations are now rapidly increasing the statistics of massive (quiescent) galaxies at much higher redshifts \citep[e.g.,][]{Carnall+24,Kawinwanichakij+26}. To our knowledge, no analysis has yet probed these scaling relations across the full relevant mass range as a function of redshift, from $z=0$ to $z \sim 3-4$, using sufficiently large samples to derive statistically robust trends. Therefore, a complete understanding of how these scaling relations evolve with redshift, environment, and galaxy type is still lacking.

Therefore further efforts are needed to explore these scaling relations as a function of redshift, and further investigations are needed to compare simulation predictions in terms of these observables with results from the literature and new future observations across redshifts.

This goal is feasible thanks to spectroscopic observations up to $z \sim 1$ from LEGA-C \citep{vanderWel+16_LEGA-C}. In the near future, surveys such as StePS-WEAVE \citep{Iovino+23_StePS-WEAVE} and StePS-4MOST \citep{Iovino+23_StePS-4MOST} will enable measurements of integrated stellar population parameters and velocity dispersions, connecting LEGA-C with local Universe results. For example, these data can be used to determine dynamical masses via Jeans modelling and infer DM fractions when combined with stellar mass estimates \citep{Tortora+09, SPIDER-VI, Tortora_Napolitano22}. With photometric and spectroscopic data at $z < 0.7$, unfortunately limited to very massive galaxies ($\mst > 10^{11} \, \rm \Msun$), we have shown that constraining the redshift evolution of size and DM fraction can provide insights into the dominant mechanisms driving their mass assembly, favouring minor mergers over major mergers. Small samples and inhomogeneous analyses exist at larger redshifts \citep[e.g.,][]{Slob+25}, and mostly focussing only on spiral galaxies \citep[e.g.,][]{Sharma+25}.

Thanks to its exquisite image quality, the spatial resolution and the wide area, {\it Euclid} will allow to discover more than 100\,000 galaxy-scale strong gravitational lenses. When combined with spectroscopic redshift and velocity dispersion measurements \citep{Collett+23_4SLSLS}, these strong lenses will make it possible to obtain the most robust and precise constraints on DM fractions and mass density slopes in thousands of strong lenses  up to $z \sim 1.5$, complementing previous analyses based on smaller samples, typically at lower redshift \citep{Gavazzi+07_SLACSIV,Koopmans+09,Auger+10_SLACSX,Treu+10,Tortora+10lensing,Sonnenfeld+13_SL2S_IV,Sonnenfeld+17_IMF,Shajib+21}.

Turning to colour and stellar population gradients, current analyses are limited to small samples of massive galaxies, typically at redshifts $z \sim 1$–2 \citep[e.g.,][]{Gargiulo+11,Gargiulo+12,Guo+11,Marian+18,Cramer+24_colour_gradients}.
{\it Euclid} will allow the exploration of various scaling relations involving structural parameters \citep{Quilley+25_Euclid_Q1},
star formation rates and stellar mass \citep{Enia+25_Euclid}. In particular, we expect to be able to study the correlation of colour gradients vs. stellar mass, and obtain constraints on stellar population gradients from the local Universe (using spatially-resolved SED fitting \citealt{Abdurrouf+25_SEDfitting_Euclid}) up to $z \sim 1.5$ \citep{Quilley+25_Euclid_Q1} with millions of galaxies at the end of Euclid operations. Instead, spatially-resolved spectroscopic data, which are less prone to systematics in the recovery of stellar population parameters with respect to colours, are particularly limited to small statistics at these redshifts \citep{Ditrani+22_gradients}, and often these analyses are only possible if these sources are gravitationally lensed \citep[e.g.][]{Akhshik+23_REQUIEM-2D}.

Tracing these scaling relations becomes increasingly challenging at higher redshifts, particularly beyond $z \sim 1-1.5$ and especially past $z \sim 2$,     which is beyond the reach of most current facilities. In addition, obtaining spatially resolved spectroscopy to accurately measure stellar population gradients and total mass profiles is considerably more difficult. Even the {\it JWST} cannot fully fill this gap, reached the needed spatial resolution and the statistics needed. To extend galaxy scaling-relation studies using both integrated and spatially resolved spectroscopy across cosmic time, especially to $z \gtrsim 2$, where the golden-mass scale may first emerge, observations with SHARP on the ELT will be essential \citep{Saracco+25_SHARP}.

In this context, this white paper outlines the scientific case for exploiting SHARP's unique capabilities to trace the emergence of the golden mass and the evolution of galaxy scaling relations.

\section{Objectives: the "SHARP" future of scaling relations at high redshift}

As stated in the previous sections, we have identified three scaling relations, connected to both stellar population parameters, dynamics and DM content of galaxies: (a) stellar population gradients, (b) DM fractions, and (c) total mass density slopes. Correlating these parameters with mass and redshift for a sample of $\sim 1500-2000$ galaxies, with $\mst \gsim 5 \times 10^{10}\, \rm \Msun$ in the redshift range $1.5-4$, can allow to constrain the contribution of different physical processes across cosmic time and for understanding the physics behind the emergence of the golden mass.

These observables can be measured through integrated or spatially-resolved spectra, both of which can be obtained with SHARP. In particular, while the DM fraction can also be constrained with fiber spectroscopy, the mass density slope and stellar population gradients are better constrained with spatially-resolved spectroscopy, thanks to SHARP unique spatial resolution. As we show in the following subsection, SHARP offers both modes, with wavelength coverage, field-of-view and spatial resolution optimal to observe galaxies at $z>1.5$ and study these scaling relations.

The current proposal relies on the measurement of various observables and is therefore connected to several science cases within the extragalactic set of proposals in this Science Book.

\subsection{Why SHARP?}\label{subsec:why_sharp}

The two complementary SHARP modes -- the AO-corrected multi-object spectrograph NEXUS and the multi-IFU spectrograph VESPER -- provide two distinct but synergistic datasets, enabling the study of stellar populations and mass distributions through both integrated and spatially resolved spectroscopy.

In NEXUS mode, approximately 30 slits can be placed within a $1.2'\times 1.2'$ field of view, with a pixel scale of 35 mas and a spectral resolution of $R\sim 2000$–6000. VESPER offers 12 integral-field units (each $1.7''\times1.5''$) with a pixel scale of 31 mas and $R\sim3000$, distributed within an area of $\sim 24'' \times 70''$. The wavelength coverage is from $0.95\, \rm \mu m$ to $2.45 \, \rm \mu m$ and $1.2\, \rm \mu m$ to $2.4 \, \rm \mu m$ for NEXUS and VESPER, respectively. 

\subsubsection{The SHARP characteristics useful for this science case}\label{subsubsec:characteristics}

SHARP has several characteristics that make it particularly well suited for the proposed observations:

\begin{itemize}
\item {\bf Wavelength coverage.}
The wavelength coverage in both observing modes includes the main absorption and emission features required to determine kinematic measurements and key stellar population parameters (SFR, age, and metallicity). In particular, this range allows the measurement of the Ca H \& K lines for galaxies in the intervals $z \in [1.5,5]$ (NEXUS) and $z \in [2,5]$ (VESPER). Moreover, age indicators such as the Balmer lines ($H_{\beta}$, $H_{\gamma}$, and $H_{\delta}$) can be observed simultaneously up to $z \sim 4$. Metallicity-sensitive features (e.g. Fe5270, Fe5335, Mgb triplet, Ca II triplet, G-band, etc.) also lie within this range, with the iron lines and Mgb triplet observable up to $z \sim 3.5$. Therefore, rotation velocity, velocity dispersion and stellar population measurements can be obtained up to $z \sim 5$ and $z \sim 3.5$–$4$, respectively. This redshift range is comparable to that shown in \Fig\ref{fig:figure2}, thus providing complementary observational constraints with respect to the star formation- and stellar-to-halo-mass relations illustrated there. Relying on the same tracers for stellar populations and kinematics as those used for local galaxies allows us to reduce biases arising from different stellar populations and facilitates comparisons across different redshifts.

In addition, for star-forming galaxies in the same redshift range, the SHARP wavelength coverage includes several strong emission lines suitable for the determination or rotation curves: [\ion{O}{II}] (up to $z \sim 5$),  H$\beta$ and [\ion{O}{III}] (up to $z \sim 3.5$), H$\alpha$ and [\ion{N}{II}] ( (up to $z \lesssim 2.5$)).

\item {\bf Spectral resolution.}
The spectral resolution ($R > 2000$) is sufficient to recover accurately the main stellar population parameters (age and metallicity), rotation velocities and velocity dispersions \citep{Iovino+23_StePS-WEAVE, Iovino+23_StePS-4MOST}.

\item {\bf Field of view.}
The multiplexing capabilities of SHARP are fully exploited in dense environments. For instance, given the number density of massive quiescent field galaxies ($\mst > 10^{10.5} \, \mathrm{M_\odot}$) at $z \sim 2$, we expect to find $\sim1$ and $\sim0.3$ such galaxies within the NEXUS and VESPER footprints, respectively \citep[e.g.,][]{Muzzin+13_mass_function,Kriek+24}. In denser environments, such as galaxy clusters at $z \gtrsim 2$ (with the number of known systems expected to increase significantly in the coming years), SHARP can probe a significant fraction of the cluster volume: areas of approximately $0.6 \times 0.6\,\mathrm{Mpc^2}$ and $0.25 \times 0.6\,\mathrm{Mpc^2}$ at $z \sim 2$ for NEXUS and VESPER, respectively. For example, 7-8 passive galaxies in the cluster at $z\sim 1.8$ studied in \cite{Newman+14} fall within the NEXUS FOV. However, placing the fibers and mIFUs on both passive and active galaxies can increase the number of allocations. This implies that a reasonable fraction of slits can be filled with targets in NEXUS mode, while a similar number of mIFUs can be allocated within two pointings in VESPER mode \citep[e.g.,][]{Newman+14,Willis+20}.  The field of view is therefore large enough to efficiently sample dense cluster regions at these high redshifts, fully optimizing SHARP’s multiplexing capability. A few control fields, targeting less dense environments and the field, would be useful to obtain homogeneous data for some representative field galaxies. These observations could be further optimized by aligning the field of view with other planned observations using SHARP.

\item {\bf Spatial resolution.}
Typical galaxies at $z \gtrsim 2$ have effective radii of $0.5$–$1 \, \mathrm{kpc}$ (corresponding to $\sim0.1''$), such that roughly half of their light falls within a $0.2''$ slit. In the mIFU mode, the light can be traced out to several effective radii. With a seeing of $\sim0.012''$ and a pixel scale of $\sim0.03''$, SHARP will be able to spatially resolve these galaxies well within the mIFU, while maximizing the S/N in slit observations. 
\end{itemize}

\subsubsection{Comparison with other intruments}

Compared to any existing or planned instrument on 8 m-class telescopes (e.g., MOONS@VLT), the large collecting area of the ELT enables SHARP in NEXUS mode to reach similar S/N with exposure times up to 20 times shorter.
It is also worth comparing SHARP with another ELT instrument, MOSAIC. Although MOSAIC benefits from a larger field of view and a higher number of fibers in its MOS mode (while the number of mIFUs is of the same order of magnitude), SHARP offers a key advantage: its broader wavelength coverage allows the measurement of velocity dispersions and stellar population parameters at higher redshifts. Specifically, velocity dispersions can be measured up to $\sim 5$ with SHARP, compared to a maximum of $\sim 3.5$ for MOSAIC, with a similar gain for stellar population diagnostics.
Furthermore, SHARP’s superior spatial resolution optimizes the signal collected within the NEXUS slits and enables galaxies to be spatially resolved in the mIFU mode, which is crucial for sampling multiple resolution elements, tracing stellar population gradients, and constraining dynamical models. Finally, with respect to {\it JWST}, SHARP offers higher spatial resolution (by more than a factor of three), and multi-object and multi-IFU capabilities.

\subsection{The survey: observational setup and exposure times}

According to the SHARP Exposure Time Calculator\footnote{\url{https://sharp.lambrate.inaf.it/}}, assuming a galaxy model with solar metallicity, an age of \(2 \, \rm Gyr\), an effective radius of \(0.1''\), and a Sérsic index \(n = 2\), the expected S/N can be estimated as follows. For an aperture of \(0.2''\) and an exposure time of 2 hours, in the NEXUS mode the average S/N is about 100 and 20 for sources with \(K = 20\) and \(K = 22\), respectively. In the VESPER mode, for a \(K = 20\) galaxy, the S/N ranges from approximately 60 at the center to 10 and 1 at \(R = R_{\rm e}\) and \(3R_{\rm e}\), respectively. When summing the signal over multiple spatial pixels at these radii, the cumulative S/N reaches about 40 and 8. For a fainter source with \(K = 22\), VESPER yields integrated S/N values of roughly 8 at \(R = R_{\rm e}\) and 3 at \(2R_{\rm e}\); increasing the exposure time to 5 hours improves these to about 13 and 5, respectively.  

Therefore, both integrated and spatially resolved measurements of stellar populations, velocities and velocity dispersions are feasible in the two observing modes with an average exposure time of 2 hours. Increasing the exposure time by 2--3 times enables probing of stellar populations and kinematics beyond one effective radius even for fainter systems.  

As discussed in \Sec\ref{subsubsec:characteristics}, also in clusters we will only be able to fill a portion of the slits at $z \gsim 1.5$--$2$, although we expect to cover a larger fraction of the cluster volume. Being conservative, we assume that $\sim 10$ slits can be allocated in NEXUS and a similar number of mIFUs in two VESPER pointings. Therefore, with 150 single pointings for NEXUS and 150 for VESPER, we can observe $\sim 1500$ and $\sim 750$ galaxies, respectively, corresponding to observing times of 300 hours each, and a total allocated time for the full proposal of $\sim 600$ hours. 

The number of sources, however, depends on the specific type of cluster observed, the redshifts, and the inclusion of both passive and active galaxies. Time allocation can also be adjusted to boost the S/N for a subsample of the faintest galaxies in a few fields, enabling robust measurements of velocities, velocity dispersions, and stellar populations beyond $1 \, \Re$.

Given the strong synergy with other SHARP programs focused on stellar populations (integrated and gradients) and dynamical analyses, collecting the required exposure times should be feasible, even for larger samples. A broader dataset would allow finer binning in stellar mass, redshift, and environment.

\subsection{Sample selection}

These numbers would provide statistically meaningful samples spanning a wide range of stellar masses and redshifts. To robustly trace the emergence of the golden mass, it is crucial that our sample selection spans across and below this characteristic scale at $z \gsim 2$. Therefore, we aim to build a mass-complete sample down to $\mst \sim 10^{10.5}\,\Msun$ at high redshift. Since one of the main scope of this contribution is the study of the golden mass itself, this lower mass limit is necessary to cover the needed mass range. Furthermore, a robust investigation of scaling relations across cosmic time must account for progenitor bias. The passive galaxies observed at high redshift represent only a fraction of the progenitors of local massive early-type galaxies; many local passive galaxies were still actively star-forming at $z \sim 2$. By simultaneously characterizing the dynamics and stellar populations of both star-forming and passive galaxies across a wide mass range, our observational strategy mitigates progenitor bias. This allows us to map evolutionary tracks more accurately and investigate how the inferred golden mass shifts depending on sample selection, completeness, and specific morphological types.

To successfully constrain the underlying physical processes (e.g., SN/AGN feedback efficiency and merger rates), robust statistics are needed to overcome the intrinsic scatter of galaxy populations. A total sample of $\sim 1500$--$2000$ galaxies with integrated and/or spatially-resolved kinematics and stellar populations is strategically sized to achieve this. To evaluate our statistical power, we adopt the $1\sigma$ scatter at fixed mass from low-redshift studies, where these scaling relations have been robustly established using statistically significant samples. Specifically, the intrinsic scatter is typically $\sim 0.1$--$0.2 \rm \, dex / \Re$ for stellar population gradients \citep{Goddard+17_gradients}, $\sim 0.2 \rm \, dex$ for the total-to-stellar mass ratio \citep{Tortora+18_KiDS_DMevol}, and $\sim 0.05 \rm \, dex$ for mass density slopes \citep{Tortora+14_DMslope}. Dividing a sample of $\sim 1500$ galaxies into 4 redshift bins and 6 mass bins yields approximately $60$ galaxies per bin. This sample size drives the standard error on the median down to $\sim 0.01$--$0.025 \rm \, dex / \Re$ for stellar population gradients, $\sim 0.025 \rm \, dex$ for the total-to-stellar mass ratio, and $\sim 0.006 \rm \, dex$ for mass density slopes. Such precision is essential to definitively distinguish subtle evolutionary trends and specific feedback signatures from measurement noise (see \Fig\ref{fig:figure1} for a qualitative visualization), thereby allowing us to accurately pinpoint the minimum associated with the golden mass.

At the moment, no definitive sample selection is possible, but a mass-limited sample at $z \gtrsim 1.5$ can be assembled from current and upcoming wide-field surveys. In particular, the \textit{Euclid} Wide and Deep Survey and the Rubin Observatory LSST will provide the bulk of the candidate systems, fully characterized in terms of stellar mass, structural parameters, and morphology, complemented by additional sources discovered and already studied with {\it HST} and {\it JWST}.

To ensure the robustness of our derived scaling relations, our survey design specifically addresses mass completeness, selection biases, and cosmic variance. First, relying on the deep near-infrared photometry from \textit{Euclid}, Rubin and \textit{JWST}, we will define a mass-completeness limit for both passive and star-forming galaxies in each redshift bin. We will apply volume corrections (e.g., the $1/V_{\rm max}$ method) to properly weight the contribution of lower-mass systems and mitigate Malmquist bias. Second, to minimize the impact of cosmic variance, which can heavily skew high-redshift galaxy mass functions and scaling relations, our allocated pointings will be distributed across multiple independent, widely separated lines of sight (e.g., sampling different \textit{Euclid} Deep Fields and several known distant clusters). This highly multiplexed, multi-field strategy ensures that our derived scaling relations reflect the universal cosmic average rather than localized over- or under-densities.

These samples will also be complemented by galaxy-scale strong gravitational lenses discovered by {\it Euclid}, with {\it JWST} extending such discoveries to higher redshift \citep{Walmsley+25_SLDE_A}. SHARP will be ideally suited to measure the stellar population properties and kinematics of both the lens and the higher-redshift source. For the lens, the combination of SHARP spectroscopy (stellar populations and kinematics) with lensing observables will enable precise determinations of the total mass \citep{Tortora+10lensing}. For the background lensed source, SHARP will further benefit from gravitational magnification, which enhances both the observed brightness and the effective spatial resolution.

By targeting both passive and star-forming systems, we will be able to assess whether the golden mass represents a universal scale affecting all galaxy types, and connect these findings with local-Universe results suggesting that star-forming and quiescent galaxies follow different star-formation efficiency trends \citep{Posti+19}.

Finally, to address the fundamental ``nature versus nurture'' dichotomy, our sample design strategically encompasses both dense galaxy clusters and isolated field regions (the latter also being integrated with independent observations). By constructing mass-matched samples of cluster and field galaxies across our targeted redshift bins ($z \sim 1.5$--$4$), we will explicitly quantify environmental differences in our scaling relations. Specifically, we will test whether the normalization, scatter, or the exact threshold of the golden mass shifts in dense environments compared to the field. Because environment-driven effects relevant to massive galaxies, such as accelerated halo quenching, enhanced rates of major and minor mergers, and tidal stripping of dark matter halos, are expected to scale differently with cosmic time compared to internal AGN/SN feedback, observing these matched samples across multiple redshift bins will allow us to empirically model the environmental evolution with redshift. This will ultimately reveal whether dense environments merely accelerate the evolutionary pathways established by internal feedback, or if they fundamentally alter the dark matter and baryonic assembly of massive galaxies.

\subsection{Methods of analysis}

To analyse the integrated and spatially resolved spectral maps of thousands of galaxies, we will use the methods detailed in the following paragraphs. A robust characterization of trends and detection of the golden mass requires a careful treatment of observational degeneracies and model-dependent systematics, as these can potentially shift, broaden, or artificially modify the observed minima in the scaling relations.

\subsubsection{Stellar populations}

Stellar populations can be characterized by comparing synthetic model predictions \citep[e.g.,][]{BC03,Maraston05,Vazdekis10,Vazdekis+12} with the main tracers of age, metallicity, and star formation rate listed in \Sec\ref{subsec:why_sharp}. Their radial variations can be investigated by dividing the data into radial bins, leading to measurements of stellar population gradients \citep{Goddard+17_MANGA}. The spatial resolution of SHARP, being three times better than {\it JWST}, allows us to trace these profiles in much greater detail, yielding more accurate and precise metallicity gradients. 

The derivation of stellar population gradients is historically plagued by the age--metallicity degeneracy, often further complicated by dust attenuation in photometric studies. However, the proposed SHARP observations directly circumvent these issues. By utilizing high-resolution spectroscopy, we will measure specific spectral indices (e.g., Balmer lines for age, and Mgb, Fe, CaT for metallicity). Because these equivalent widths are measured relative to the local continuum, they are intrinsically robust against dust reddening, which alters the broadband slope but leaves narrow spectral features unaffected. Most importantly, SHARP's broad wavelength coverage ($0.95$--$2.45\,\rm \mu m$) allows us to track these distinct age and metallicity features simultaneously, definitively breaking the degeneracies that could otherwise artificially steepen or flatten the measured gradients.

\subsubsection{Dynamical and kinematical analysis}

Dynamics, and therefore DM fraction and mass density slopes can be obtained using both the modes: slits and mIFUs. By measuring velocity dispersions and rotation velocities in slits, we can constrain the central total mass and DM fraction within fixed radii (e.g. the effective radius), and  mass density slopes, by making some assumptions of the galaxy models \citep{Tortora+09,SPIDER-VI,Tortora+14_DMslope,Tortora+18_KiDS_DMevol}. With observations in mIFU, for a smaller sample of galaxies, we will measure spatially-resolved profiles for velocity and velocity dispersion. This provides stronger constraints on the DM fraction (as a function of radius), the total and DM density slopes. 

Two source of systematics exist in this analysis. First, the inferred DM fractions and mass density slopes may be sensitive to the adopted dynamical model and the assumed DM halo profile (e.g. \citealt{SPIDER-VI,Tortora+14_DMslope}). However, as pointed out in \cite{Busillo+25_CASCO-II}, the Jeans Anisotropic Modelling (JAM) method applied to spatially resolved spectroscopy from \cite{Zhu+23_MangadynpopI} shows that differences in DM fractions and mass density slopes are smaller than the uncertainties across different model setups. Even if systematic biases in the adopted mass model may shift the absolute values of the DM fraction and the total mass density slope, the overall trends remain largely unaffected, and the presence of the U-shaped minimum is robust across the galaxy population, even though its exact mass scale may shift by a few tenths of a dex due to modelling assumptions. Second, the assumption of a universal Initial Mass Function (IMF) represents a major source of systematic uncertainty. Extensive literature has shown that the IMF is not universal, becoming increasingly bottom-heavy in more massive early-type galaxies \citep[e.g.,][]{Spiniello+12,TRN13_SPIDER_IMF,LaBarbera+13_SPIDERVIII_IMF}. For direct comparisons with cosmological simulations, the correlation between DM fraction and stellar mass is generally investigated under the assumption of a universal Milky Way-like IMF. Adopting a variable IMF would likely make this scaling relation shallower; however, the minimum is expected to persist in both the correlations \citep{Tortora+14_DMslope,TLBN16_IMF_dwarfs,Tortora+19_LTGs_DM_and_slopes}. Fortunately, SHARP will mitigate this source of bias. Its NIR spectral coverage will enable measurements of IMF-sensitive absorption features, which, combined with spatially resolved dynamical modelling, may allow the IMF to be constrained directly and the intrinsic shape of the scaling relations to be determined.

These analyses can be performed through Jeans modeling \citep{Tortora+09,SPIDER-VI,Tortora_Napolitano22} or more sophisticated approaches such as Schwarzschild orbit-superposition modeling \citep{Cappellari+06,ThomasJ+07}. Rotation curves are used for purely rotating systems. In the era of big data, machine learning techniques can facilitate and accelerate these analyses or provide alternative methods to determine such quantities, for example by learning the correlations between measured observables, such as stellar mass and velocity dispersion, with DM fraction and mass density slope \citep{vonMarttens+22,Wu+24}, or by assisting in the analysis of spatially-resolved spectra. If the target is the deflector in a gravitational lensing system, the lensing features can complement dynamical observables, yielding even more precise constraints on the total mass and mass density profiles \citep{Tortora+10lensing,Auger+10_SLACSX}.

Also in this case, the high spatial resolution of SHARP is helping, providing the necessary resolution to trace in detail the dynamical maps and distinguish between the contribution from velocity and velocity dispersion.

A mass model must be chosen in the Jeans modeling or in the modelling of the rotation velocity, either assuming a single component for the galaxy (e.g., SIS or constant-M/L profile) or a two-component model that separates the stellar and DM contributions, with the latter often taken from N-body simulations \citep{Tortora_Napolitano22}. Systematics arising from the choice of model, the inclusion of a central black hole, or anisotropy need to be carefully analysed, and high spatial resolution can help mitigate these uncertainties. IMF non-universality and its potential radial gradient must also be considered \citep{LaBarbera+13_SPIDERVIII_IMF, Spiniello+14}; the IMF can be treated as a free parameter in the dynamical analysis and dynamically constrained using data from both slits and mIFU observations \citep[e.g.,][]{Cappellari+12,TRN13_SPIDER_IMF}.

Crucially, within the relevant stellar mass range, a fundamental kinematic transition in early-type galaxies is found, precisely around $\sim 10^{11}\, \Msun$ \citep{Emsellem+07_SAURON_IX, Cappellari+13_ATLAS3D_XX}. The early-type population transitions from rotation-supported ``fast rotators'', which dominate at lower masses, to dispersion-dominated ``slow rotators'' at the high-mass end, largely driven by dry major mergers. Understanding when and how this kinematic dichotomy emerges in the early Universe remains an open question, as does whether this mass scale coincides with the golden mass. SHARP's VESPER mode, with its exceptional spatial resolution and mIFU capabilities, is uniquely positioned to map the spatially-resolved kinematics (e.g., the stellar specific angular momentum) of these high-redshift systems, directly linking the golden mass to the dynamical assembly of galaxies.

\section{Summary and conclusions}

In summary, both observations and simulations are steadily advancing our understanding of galaxy formation and evolution across cosmic time. Astronomers have measured and continue to measure scaling relations among structural and stellar parameters, as well as total and DM fractions, as functions of mass and redshift. With ongoing and upcoming spectroscopic (WEAVE and 4MOST) and imaging (Euclid and LSST) surveys, our knowledge of the Universe is rapidly expanding. Nevertheless, the earliest epochs remain largely unexplored, even though {\it JWST} has recently begun to unveil the properties of hundreds of very high-redshift galaxies.

Yet, many key questions remain unanswered. Which physical processes shape the observed scaling relations, and to what extent? Why do these relations exhibit characteristic mass scales, such as the golden mass, and how do they manifest? How do these relations and characteristic masses vary with galaxy type and cosmic epoch?

SHARP@ELT, with its unique combination of wavelength coverage, spatial and spectral resolution, and multiplexing capabilities, will enable us to extend spectroscopic studies to higher redshifts than currently possible with existing instruments, even at improved spatial resolution. In this paper, we have identified specific correlations of three observables with stellar mass: (a) stellar population gradients, (b) DM fraction, and (c) total mass density slope. We have explained how their combined analysis as a function of redshift can provide powerful and independent constraints on the differential role of galaxy formation processes across different masses, as well as on the emergence of the golden mass.

These results will complement, among the others, those obtained from {\it Euclid} and LSST imaging, {\it JWST} imaging and spectroscopy, as well as Euclid’s gravitational lensing-based mass measurements.

With a total of $\sim 600$ hours of observations, we plan to observe at least 1500 and 750 galaxies in the NEXUS and VESPER modes, respectively. This unique statistical sample of galaxies in the redshift range $z \sim [1.5,4]$ will allow us to derive these correlations with high precision, trace the evolution of the golden mass as a function of redshift, and disentangle the roles of feedback processes and environment. We have also illustrated examples of how comparisons with cosmological simulations can already provide insights \citep{Busillo+23_CASCO-I,Busillo+25_CASCO-II,Tortora+25_CASCOIII}, and future large-volume, higher-resolution simulation suites will enable even more detailed comparisons and physical insights once SHARP begins observing with the ELT.

\section*{Acknowledgements}
CT acknowledges the INAF grant 2022 LEMON. FLB acknowledges support from INAF 2023 RSN1
Minigrant 1.05.23.04.01. FLB acknowledges support from
the Fundación Occident and the Instituto de Astrofísica de
Canarias under the Visiting Researcher Programme 2022-
2025 agreed between both institutions. 
RR acknowledges financial support through INAF-WEAVE StePS funds and through PRIN-MIUR 2020SKSTHZ. The SHARP team acknowledge support from the INAF Techno-Grant 2022, SHARP - 1.05.12.02.01, and INAF Large-Grant 2024, SHARP -
1.05.24.01.01. The SHARP team acknowledges support by Bando Ricerca Fondamentale INAF 2022,
Techno-Grant "SHARP" - 1.05.12.02.01 and Bando Ricerca Fondamentale INAF 2024,
Large-Grant "SHARP" - 1.05.24.01.01. 

\clearpage 






\bibliographystyle{aa}

\bibliography{myrefs}



\end{document}